\documentclass[12pt,authoryear]{article}
\usepackage[paperheight=11.7in,paperwidth=8.3in,margin=1in]{geometry}

\usepackage[utf8]{inputenc}
\usepackage{amsmath}
\usepackage{amssymb}
\usepackage{amsfonts}
\usepackage{bm}

\usepackage{appendix}
\usepackage{enumerate}
\usepackage{enumitem}
\usepackage{nicefrac}
\usepackage{array}
\usepackage{calc}
\usepackage{flafter}
\usepackage{esdiff}
\usepackage{mathabx}
\usepackage{textcomp}
\usepackage{color}		
\usepackage{graphicx}
\usepackage{amscd}
\usepackage{mathrsfs}
\usepackage{enumerate}
\usepackage{mdwlist}
\usepackage[round]{natbib}
\usepackage{authblk}
\usepackage[colorlinks=true,linkcolor=blue,citecolor=blue,urlcolor=blue]{hyperref}
\usepackage[export]{adjustbox}

\usepackage{caption}
\usepackage{subcaption}

\hypersetup{
    colorlinks,
    citecolor=black,
    filecolor=black,
    linkcolor=black,
    urlcolor=black
}

\makeatletter
\def\BState{\State\hskip-\ALG@thistlm}
\makeatother

\makeatletter
\def\step{%
   \@ifnextchar[ \@myitem{\@noitemargtrue\@myitem[\@itemlabel]}}
\def\@myitem[#1]{\item[#1]\mbox{}}
\makeatother

\newcommand\R{\mathbb{R}}

\newcommand\E{\mathbb{E}}

\newcommand{\calG}{\mathcal{G}}

\newcommand{\calV}{\mathcal{V}}

\newcommand{\ra}{\rightarrow}
\newcommand{\la}{\leftarrow}

\renewcommand\epsilon{\varepsilon}

\newcommand{\Exp}{\text{Exp}}
\newcommand{\Log}{\text{Log}}

\DeclareMathOperator*{\argmax}{arg\,max}
\DeclareMathOperator*{\argmin}{arg\,min}

\DeclareMathOperator*{\arginf}{arg\,inf}

\graphicspath{{../}}

\title{Functional random effects modeling of brain shape and connectivity}
\author[1]{Eardi Lila\thanks{elila@uw.edu}}
\author[2]{John A. D. Aston\thanks{j.aston@statslab.cam.ac.uk}}
\affil[1]{Department of Biostatistics, University of Washington}
\affil[2]{Statistical Laboratory, DPMMS, University of Cambridge}
\date{}

\begin{document}
\maketitle
\begin{abstract}
\noindent We present a statistical framework that jointly models brain shape and functional connectivity, which are two complex aspects of the brain that have been classically studied independently. We adopt a Riemannian modeling approach to account for the non-Euclidean geometry of the space of shapes and the space of connectivity that constrains trajectories of co-variation to be valid statistical estimates. In order to disentangle genetic sources of variability from those driven by unique environmental factors, we embed a functional random effects model in the Riemannian framework. We apply the proposed model to the Human Connectome Project dataset to explore spontaneous co-variation between brain shape and connectivity in young healthy individuals.
\end{abstract}

\section{Introduction}\label{sec:intro}
Human brains differ in their structural and functional organization \citep{Gilmore2018}. While there is a long history of trying to relate either structural or functional brain features to human aspects, such as behavioral and cognitive variables \cite[for recent examples, see, e.g.,][]{Xia2018, Zhang2019}, more recently, increasing attention has been drawn to the problem of understanding how brain structure and function are related to each other \citep{Bullmore2009}.

\begin{figure}[!htb]
\centering
\includegraphics[width=.9\textwidth]{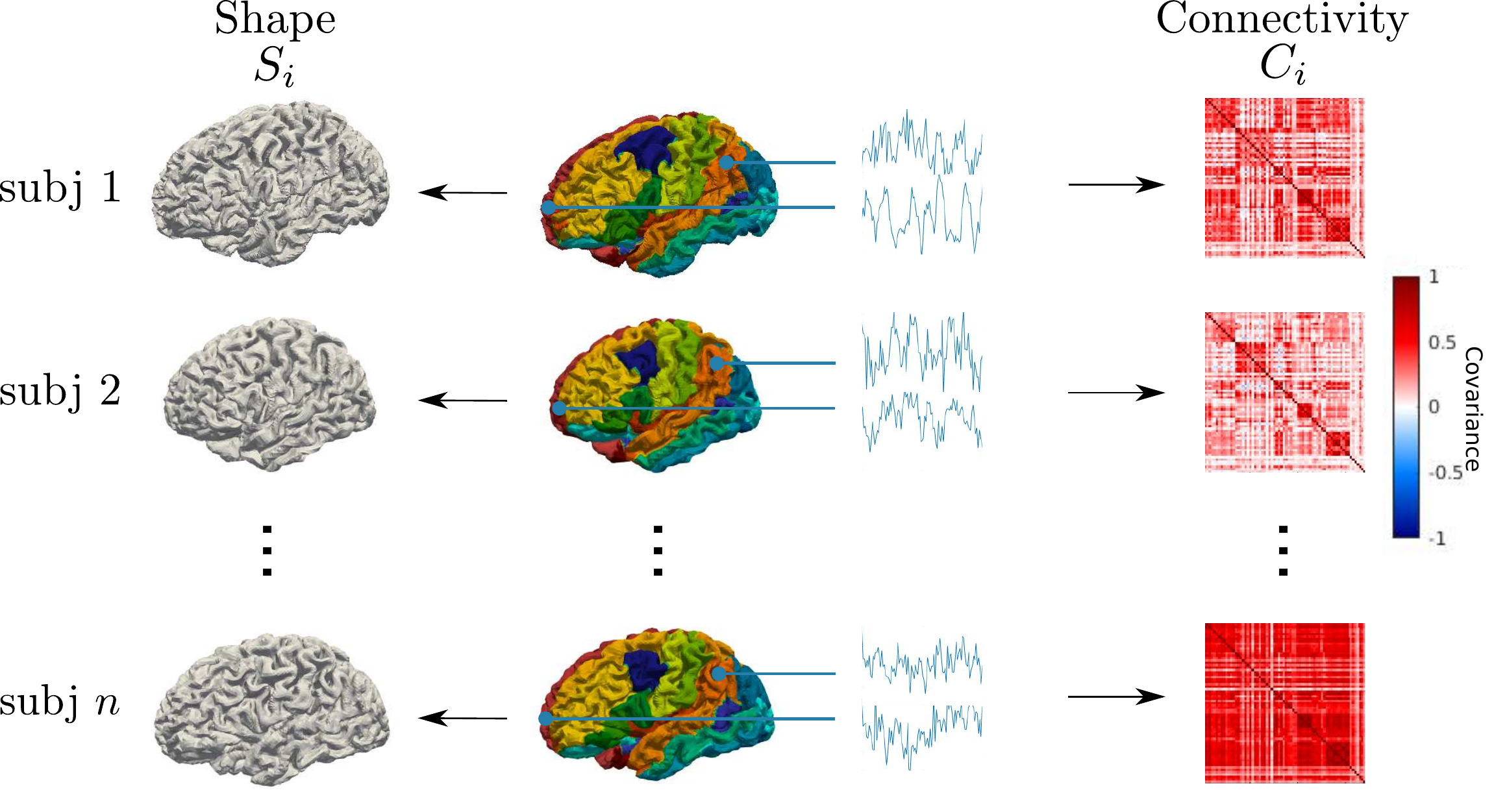}
\caption[]{In the central panel, we show the subject-specific surfaces encoding the geometry of the cerebral cortex, as reconstructed from the MRI scans. Moreover, we can see the fMRI time-series, describing the neuronal activity of a dense set of 64K points on the cerebral cortex. The color map on the brain surfaces describes a parcellation atlas, which defines 68 regions of the brain in correspondence across subjects. Within each region, an average time-series is computed. These are then used to compute the $68 \times 68$ covariance matrices on the right panel, describing the functional connectivity of each subject. On the left panel, a representation of only the shape of the brain surfaces. The shapes on the left panel and the covariances on the right panel are the object-data of our statistical analysis.}
\label{fig:data}
\end{figure}

In this work, we introduce a statistical framework that allows us to estimate patterns of co-variation between brain structure and function, while disentangling co-variation due to genetic and environmental factors. We describe the brain structural organization of an individual with a surface encoding the \textit{brain shape}, that is the geometry of the highly convoluted outermost layer of the brain, called the cerebral cortex. We describe the brain functional organization of an individual with a network that has spatial nodes located on the cerebral cortex. The strength of the network edges is estimated by a measure of pairwise statistical dependence (e.g., correlation) between the neuronal activity associated with the network nodes. The estimated network is a representation of the subject's \textit{brain functional connectivity}. Figure~\ref{fig:data} provides an illustration of this setting.

We apply the proposed methodology to $1003$ young adults in the Human Connectome Project (HCP) dataset \citep{Glasser2013} with the aim of exploring spontaneous modes of genetically-driven and environmentally-driven co-variation in the brain structure and function of healthy individuals.

From a methodological perspective, our work can be contextualized within the Object Data Analysis framework \citep{Marron2014} as both shapes and connectivity networks are complex data objects, living on functional non‐Euclidean spaces, where classical Functional Data Analysis approaches \citep{Ramsay2005} fail to preserve the geometry of these spaces. In order to enforce physiologically valid shape trajectory estimates, we represent brain shapes through diffeomorphic deformation functions of the ambient space. We represent brain connectivity by means of covariance functions, which must be non-negative definite. Our approach tackles diffeomorphic constraints and non-negative definiteness constraints in a Riemannian framework, i.e., by tangent space mapping through Riemannian logarithmic maps.

Within the proposed Riemannian modeling framework, we define a multi-variable trait \textit{variance component model} that exploits the relatedness structure among individuals to disentangle co-variation in shape and connectivity that is due to genetic sources and environmental sources. The proposed model can be regarded as an extension of the classical single-trait and bivariate-trait variance component models in \cite{Amos1994,Almasy1997} and is formulated as a multivariate mixed effects model on the Karhunen–Lo\`eve basis coefficients of the tangent space coordinates. Related models have already been employed in the neuroimaging literature to estimate the impact of genes and environment on structural brain development \citep{Lenroot2008} or to model subject-specific heterogeneity in functional connectivity \citep{Fiecas2017}. However, such studies tend to focus on either structural or functional features.

Mixed effects models have been successfully extended to the setting of functional data in a linear space, to account for non-parametric fixed and random effects \citep[see, e.g.,][]{Shi1996,Guo2002,Wu2002,Qin2005,Morris2006,Chen2008,
Zhou2008,Reimherr2016,Liu2017,Scheipl2015}. 
Functional models that incorporate genetic information, without explicitly relying on a mixed effects model have also been formulated. For instance, \cite{Kirkpatrick1989} introduce a model to separate genetic functional traits and \cite{Luo2019} propose a model that is able to dissect genetic and environmental effects of functional data in a twin study design. The proposed model applies to linear functional data and is formulated as a functional structural equation model. An extension to functional data over two-dimensional domains and living in a linear function space, such as cortical thickness data mapped onto a spherical domain, has been proposed in \cite{Risk2019}. In the non-Euclidean framework of our analysis, the variance component model approach allows for more flexible relatedness structure among individuals, possibly estimated from Single Nucleotide Polymorphism (SNP) data \citep[see, e.g.,][]{Dahl2016}.

The HCP dataset, which motivates this work, includes Magnetic Resonance Imaging (MRI) and resting-state functional MRI (fMRI) scans. The MRI scans are used to reconstruct surface models of the cerebral cortex geometry. The time-variant fMRI signals are used to estimate a spatial covariance structure on the cerebral cortex, describing how the different parts of the cerebral cortex co-activate in time, namely an estimate of the functional connectivity. An illustration of the MRI and fMRI components of the data is provided in Figure~\ref{fig:data}. Moreover, the pedigree of the HCP cohort is available and includes monozygotic twins, dizygotic twins, full siblings, half-siblings, and unrelated individuals. This family structure allows the variance component model to disentangle genetic and environmental co-variation between brain shape and connectivity.

\subsection*{Statistical analysis of shapes and covariances}

In the neuroimaging literature, brain shape is usually modeled using a few descriptors of shape, such as cortical volume or the area of pre-defined sets of brain regions \citep{Im2008a,Hazlett2017}. In the statistical literature, a non-exhaustive list of shape analysis methodologies based on discrete representations includes landmark-based shape representations \citep[see, e.g.,][]{Dryden2016}, skeletal shape representations \citep{Pizer2013}, dihedral angles representations \citep{Eltzner2018} and projective shape spaces \citep{Mardia2005}. In the continuous setting of curves and surfaces, global parametrizing functions have been adopted to represent these objects.

Representing curves and surfaces with their parametrizing functions, equipped with an $L^2$ norm, leads to unnatural trajectories in the space of shapes. Instead, a successful approach consists of equipping the space of parametrizing functions with an Elastic Riemannian metric and defining parametrization-invariant representations. The resulting space and associated metric lead to naturally looking geodesic trajectories in the space of curves \citep{Kurtek2012,Su2014} and surfaces \citep{Kurtek2011,Jermyn2012,Jermyn2017}.

Of particular importance to this work is a class of representation models for surfaces that do not require the computation of parametrizing functions. These represent surfaces with diffeomorphic deformation functions of the ambient space $\R^3$ \citep[see, e.g.,][]{Vaillant2004,Charon2014,Arguillere2016,Younes2010}. Such an approach is well suited to the neuroimaging setting because constraining the deformation functions to be diffeomorphic results in a shape space that contains anatomically plausible shapes and excludes, for instance, self-intersecting surfaces. Statistical analysis can then be performed on the non-linear manifold of diffeomorphic functions by exploiting a tangent space expansion to find a linear representation of the data.

In a similar spirit to shape analysis, the statistical analysis of samples that are covariances also involves a non-Euclidean type analysis. In fact, the neuroimaging community has often approached the problem by performing multivariate analysis on vectorizations of the covariances \citep{Smith2015,Xia2018}. However, such an approach fails to guarantee that linear extrapolations of the data belong to the space of covariances, i.e., that they are positive semi-definite objects. In other words, a signal with the estimated extrapolated `covariance' may not exist. Such an issue can be overcome by defining an appropriate Riemannian metric on the space of covariances.

To this purpose, different metrics have been proposed. For instance, \cite{Pennec2006} introduce an affine invariant Riemannian metric, while \cite{Arsigny2006} introduce a log-Euclidean metric based on the matrix exponential and matrix logarithm functions. \cite{Dryden2009} introduce a metric that can deal with rank deficient covariance matrices, and its extension to covariance operators has been proposed in  \cite{Pigoli2014}. As shown in \cite{Dryden2009}, different metrics lead to different geodesic trajectories in the space of covariances. While these trajectories are easy to visualize for lower dimensional covariances, in our high-dimensional setting, these differences are more difficult to appreciate. Therefore, our choice of the metric is mostly driven by computational efficiency arguments, and in particular by closed-form solutions of the geodesic mean. Statistical analysis can then be performed on tangent space projections of the covariances, computed through the Riemannian logarithmic map, which offer a convenient linear parametrization. We then use the tangent coordinates in the space of shapes and the space of covariances to find maximally associated modes of variation in shape and connectivity while decomposing the genetic and unique environmental variance contributions.

The rest of the paper is organized as follows. In Section~\ref{sec:model_sec}, we introduce the Riemannian modeling framework and the variance component model. In Section~\ref{sec:estimation} we present the implementation details of the proposed model. We then apply the proposed model to the HCP dataset and present the results in Section~\ref{sec:application}. We finally give some concluding remarks in Section~\ref{sec:conclusion}. The simulations validating the variance component model are postponed to the appendix.

\section{Mathematical description of the model}\label{sec:model_sec}
Consider a sample of $n$ pairs of observations $\{(S_i,C_i): i=1,\ldots,n\}$. Here, $(S_i \subset \R^3)$ are two-dimensional surfaces, embedded in $\R^3$, representing the cerebral cortex geometries. The functions $(C_i:S_i \times S_i \ra \R)$ are covariance functions representing the associated functional connectivity on the subject's cerebral cortex.  Moreover, we assume statistical relatedness between the $n$ samples; in our application this is due to family-based genetic associations. The aim of this section is to introduce the proposed statistical framework for the analysis of the objects $\{(S_i,C_i): i=1,\ldots,n\}$.

As previously mentioned, both the space of brain geometries and that of brain connectivity are non-Euclidean spaces, introducing additional challenges in the definition and estimation of the co-variation structure. In Section~\ref{sec:model}, we first give a brief conceptual description of the Riemannian approach to modeling shape and connectivity spaces, and then follow by introducing the variance component model. We detail our choices of the representation models and metrics, for the shape and connectivity spaces, in Section~\ref{sec:model_shape} and Section~\ref{sec:model_cov}, respectively.

\begin{figure}[!htb]
\centering
\includegraphics[width=1\textwidth]{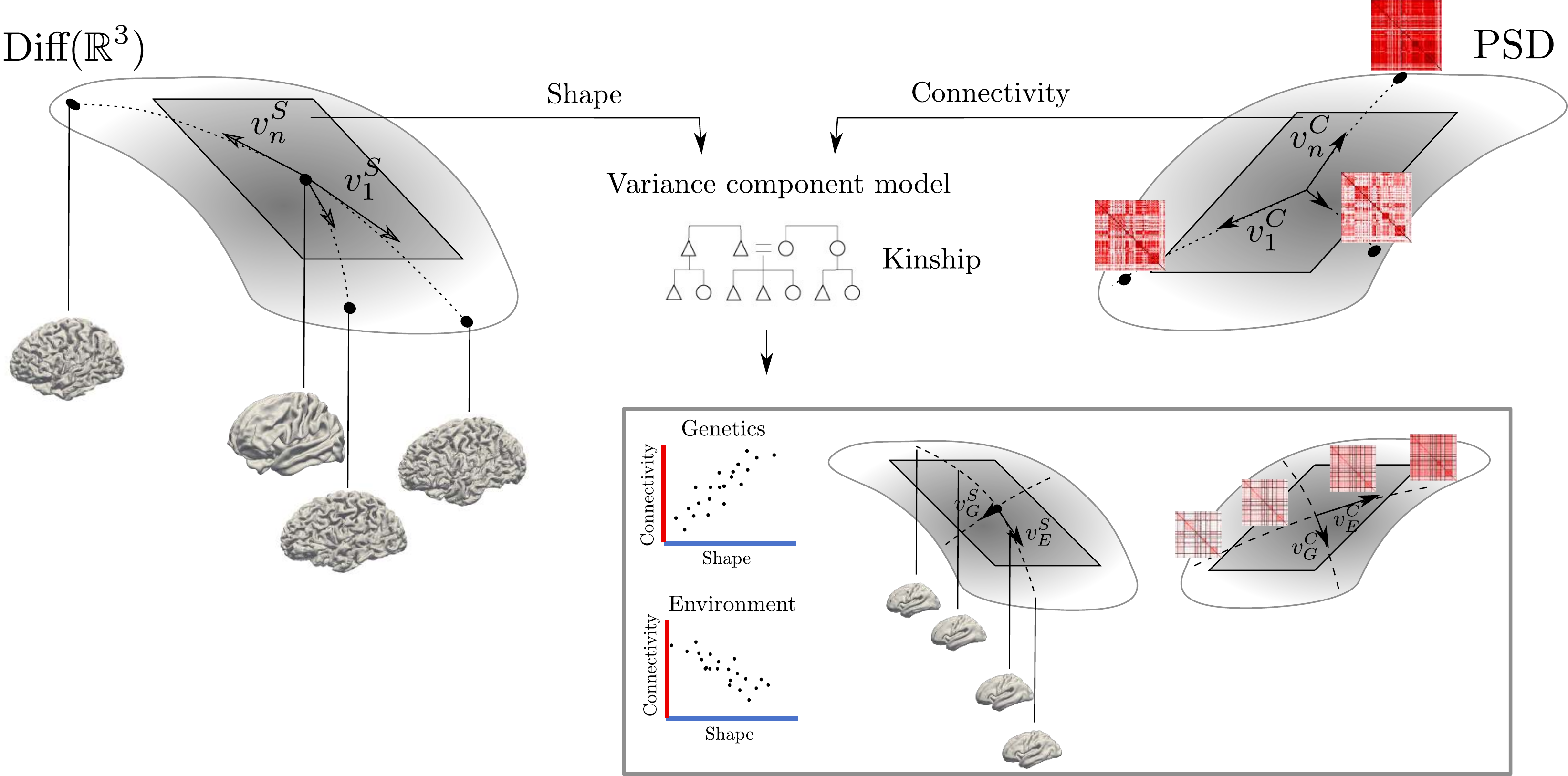}
\caption[]{This is an illustration of the proposed statistical analysis framework. We represent shapes with diffeomorphic deformations of the ambient space. The depiction of the space of diffeomorphic functions and that of covariances highlights their non-Euclidean structure. We rely on a tangent space expansion to derive linear parametrizations of these non-Euclidean spaces. The linear tangent coordinates of shape $(v_i^S)$ and connectivity $(v_i^C)$ are then jointly used to define a variance component model that exploits the kinship structure among the samples to estimate pairs of tangent coordinates that are highly correlated due to genetic factors $(v_G^S,v_G^C)$ or environmental factors $(v_E^S,v_E^C)$. The shape and covariance non-linear trajectories associated with the estimated pairs of tangent coordinates can finally be computed to display the results in terms of elements of the shape and connectivity spaces.}
\label{fig:analysis}
\end{figure}

\subsection{Functional random effects modeling of shape and connectivity}\label{sec:model}
Let $S_0 \subset \R^3$ be a template surface. We assume there exists a one-to-one correspondence between each of the points on $S_0$ and those on a surface $S_i$. The role of the template is two-fold here. The template is a surface representing an ``average'' geometric shape of the population that allows us to model shapes as functions that are $\R^3$ deformations of the reference template. Moreover, the template plays the role of a common reference domain where the subject-specific covariance functions can be mapped.

For a fixed template, we represent each surface $S_i$ with an associated deformation function $\gamma_i: \R^3 \ra \R^3$ such that $\gamma_i(S_0) = S_i$. These deformations are diffeomorphic functions, i.e., they are smooth one-to-one functions with smooth inverse. The space of deformations is formally equipped with a Riemannian metric and the diffeomorphic functions are projected onto the tangent space centered at the identity function. A set of tangent space coordinates $v_1^S, \ldots, v_n^S$ is then used to represent the surfaces $S_1, \ldots, S_n$. We assume that each function $v_i^S$ can be expressed in terms of a common basis expansion
\[
v_i^S = \sum_{j=1}^\infty A_{i,j}^S \psi_j^S,
\]
where $ \psi_j^S$ is the $j$th basis element and $A_{i,j}^S$ is the coefficient of the $i$th sample associated with the $j$th basis.

The covariance functions $(C_i)$ also belong to a non-Euclidean space, which is the cone of positive semi-definite covariance functions. Moreover, they are defined on sample-specific spatial domains $(S_i)$. The previously defined deformation functions $(\gamma_i)$ can be used to map a covariance $C_i$ onto the template $S_0$, defining $C^0_i(x,y) := C_i(\gamma_i^{-1}(x),\gamma_i^{-1}(y))$, with $x,y \in S_0$. This leads to a set of `spatially normalized' covariance functions $C^0_i:S_0 \times S_0 \ra \R$. 
As detailed in Section~\ref{sec:model_cov}, prior to the definition of the Riemannian metric we reduce the covariance functions into finite-dimensional positive-definite covariance matrices. We omit the details here to keep the notation simple, however, it should be noted that this step has implications on the class of metrics that we will be able to adopt \citep[see, e.g.,][for a discussion]{Pigoli2014}. The data $(C^0_i)$ are then projected onto the tangent space centered at the geodesic mean and the associated tangent space coordinates $v_1^C, \ldots, v_n^C$ are used to represent $C^0_1,\ldots,C^0_n$. As for the shape tangent coordinates, we assume $(v_i^C)$ can be expressed in terms of a common basis expansion
\[
v_i^C = \sum_{j} A_{i,j}^C \psi_j^C,
\]
with $\psi_j^C$ the $j$th basis element and $A_{i,j}^C$ the coefficient of the $i$th sample associated with the $j$th basis.

A simple approach to controlling for a set of known confounding variables $(z_i \in \R^l)$ consists of modeling the coefficients $\E\left[A_{i,j}^S|z_i\right]$ and $\E\left[A_{i,j}^C|z_i\right]$ through a regression analysis. The conditional expected values $\E\left[A_{i,j}^S|z_i\right]$ and $\E\left[A_{i,j}^C|z_i\right]$ are related to $\E \left[v_i^S | z_i\right]$ and $\E \left[v_i^C | z_i\right]$ by the following equations:
\begin{align*}
\E \left[v_i^{S} | z_i\right] = \sum_{j} \E \left[A_{i,j}^S | z_i \right] \psi_j^S, \qquad
\E \left[v_i^{C}| z_i\right] = \sum_{j} \E \left[A_{i,j}^C | z_i \right]\psi_j^C
\end{align*}
The estimated effects of the confounders can then be removed from the tangent space coordinates.

In practice, we choose appropriate truncation levels $p_S$ and $p_C$, and rely on the finite-dimensional approximations
\[
v_i^S \approx \sum_{j=1}^{p_S} A_{i,j}^S \psi_j^S, \qquad v_i^C \approx \sum_{j=1}^{p_C}  A_{i,j}^C \psi_j^C.
\]
While any basis in the space of the tangent coordinates is a valid choice, in our application we adopt a principal component basis, due to its well known best linear approximation property.

Before introducing our variance component model, we briefly recall the matrix normal distribution $\operatorname{MVN}$, which generalizes the multivariate normal distribution to matrix-valued random variables. A $n \times p$ random matrix $R$ has a matrix normal distribution $\operatorname{MVN}(M, U, V)$ if and only if $\operatorname{vec}(R)$ has a multivariate normal $\mathcal{N}(\operatorname{vec}(M), V \otimes U)$, where $\operatorname{vec}$ is the column-wise vectorization operator, and $\otimes$ denotes the Kronecker product. The matrix normal distribution is characterized by three parameters that are a $n\times p$ mean matrix $M$, a $n\times n$ matrix $U$ (modeling the covariance structure between the rows of the random matrix) and a $p\times p$ matrix $V$ (modeling the covariance structure between the columns of the random matrix).

Consider now the $n \times p_S$ coefficient matrix $(A^S)_{ij} = A_{i,j}^S$ and the $n \times p_C$ coefficient matrix $(A^C)_{i,j} = A_{i,j}^C$. We propose a joint model for shape and connectivity in terms of their $n \times (p_C+p_S)$ coefficient matrix $A$ that contains both the features described in $A^S$ and $A^C$, i.e.,
\begin{align*}
A := &\begin{bmatrix}
A^S, A^C
\end{bmatrix}.
\end{align*}

If at this stage we were interested in understanding co-variation between brain shape and connectivity, we could, for instance,  perform canonical correlation analysis on the scores matrices $A^S$ and $A^C$, or alternatively, perform an angle-based joint and individual variation analysis \citep{Feng2018,Carmichael2019}. In fact, several related approaches have been proposed to integrate structural and functional neuroimaging data \citep[see, e.g.,][]{Franco2008, Sui2011, Xue2015}. However, the estimated joint variation components would be an average of the co-variation that is due to genetic and environmental factors. Therefore, our next step is defining a model that separates genetic and environmental variability.

Let $K_n$ be a $n \times n$ matrix of relatedness coefficients, that is, a correlation structure, among the $n$ subjects, encoding genetic relatedness. We assume this is known. In practice, it can be estimated from a pedigree or from genomic data \citep{Lange2002,Kang2010,Wang2017b}. Let $I_n$ be the $n \times n$ identity matrix, which as opposed to $K_n$, encodes `unrelatedness' between subjects. Let the $(p_S+p_C)\times (p_S+p_C)$ matrices $\Sigma_G$ and $\Sigma_E$ denote respectively the unknown genetic and environmental covariance structure across the columns of ${A}$. We model $A$ as
\begin{equation}\label{eq:multi-trait}
\begin{aligned}
&A \,| \,G,E  =
XB + G + E\\
&G \sim \operatorname{MVN}(0_{n\times (p_S+p_C)}, K_n, \Sigma_G),\\
&E \sim \operatorname{MVN}(0_{n\times (p_S+p_C)}, I_n, \Sigma_E),
\end{aligned}
\end{equation}
where $G$ and $E$ are independent, $X$ denotes a $n \times s$ fixed design matrix and $B$ a $s \times (p_S+p_C)$ matrix of fixed effects (i.e. not random). In our final application, we do not include a fixed effect term and focus only on the variance component terms $G$ and $E$.

The proposed model exploits a known covariance structure $K_n$, across the samples, to additively decompose the covariance structure across the traits into two components, that are $\Sigma_G$ and $\Sigma_E$. In our application, $K_n$ is chosen to reflect the pairwise family relatedness across samples. Therefore, we refer to $\Sigma_G$ as the covariance component of the traits that is due to additive genetic contributions, while we refer to $\Sigma_E$ as the covariance structure that is due to unique environmental contributions. In practice, $\Sigma_G$ and $\Sigma_E$ are estimated with a Restricted Maximum Likelihood (REML) approach, as detailed in Section~\ref{sec:mixed_effect_impl}.

The model proposed is a multi-variable trait extension of the polygenic quantitative trait variance component model \citep[][]{Amos1994,Almasy1997}. This is also related to multivariate linear mixed models applied in genome-wide association studies \citep{Zhou2014b, Dahl2016}. In our model, the multivariate traits are tangent space descriptors of brain shape and connectivity.

The matrices $\Sigma_G$ and $\Sigma_E$ can be written as
\[
\Sigma_G =
\left[
\begin{array}{c c}
\Sigma_G^{S,S} &  \Sigma_G^{S,C}\\
\Sigma_G^{C,S} &  \Sigma_G^{C,C}
\end{array}
\right], \qquad
\Sigma_E =
\left[
\begin{array}{c c}
\Sigma_E^{S,S} &  \Sigma_E^{S,C}\\
\Sigma_E^{C,S} &  \Sigma_E^{C,C}
\end{array}
\right],
\]
with $\Sigma_G^{S,S}$ a $p_S \times p_S$ matrix, $\Sigma_G^{C,C}$ a $p_C \times p_C$ matrix, $\Sigma_G^{S,C}$ a $p_S \times p_C$ matrix, and $\Sigma_G^{C,S}$ a $p_C \times p_S$. The environmental components are defined similarly. These matrices represent the covariances between and within low-dimensional representations of the tangent space coordinates of shape and connectivity. In this form, they are not themselves very informative. Therefore, we propose to look at maximally correlated genetic and environmental modes of co-variation between the geometry of the cerebral cortex and associated connectivity by computing
\[
(\theta_G^S, \theta_G^C) = \left\{ \argmax_{\theta,\eta} \theta^T \Sigma_G^{S,C} \eta : \theta^T\Sigma_G^{S,S}\theta = 1, \eta^T \Sigma_G^{C,C} \eta = 1 \right\}
\]
and
\[
(\theta_E^S, \theta_E^C) = \left\{\argmax_{\theta,\eta} \theta^T \Sigma_E^{S,C} \eta : \theta^T\Sigma_E^{S,S}\theta = 1, \eta^T \Sigma_E^{C,C} \eta = 1 \right\},
\]
where $\theta_G^S, \theta_E^S$ are $p_S$-dimensional vectors, while $\theta_G^C, \theta_E^C$ are $p_C$-dimensional vectors. Constraints of the type $\theta^T \theta = 1, \eta^T \eta = 1$ are also valid choices. The pairs $(\theta_G^S, \theta_G^C)$ and $(\theta_E^S, \theta_E^C)$ represent the maximally correlated modes of co-variation in shape and connectivity that are due to genetic and non-genetic factors. Subsequent modes of co-variation can also be estimated by maximizing the same objective functions while imposing orthogonality constraints with respect to the previously computed components.

We then compute the tangent space coordinates associated with these modes of co-variation, i.e.,
\begin{align*}
\left( v^S_{G},v_G^C \right) := \left( \sum_{j=1}^{p_S} \theta_{j,G}^S \psi_j^S, \sum_{j=1}^{p_C} \theta_{j,G}^C \psi_j^C \right), \qquad
\left(v^S_{E}, v_E^C \right) := \left( \sum_{j=1}^{p_S} \theta_{j,E}^S \psi_j^S, \sum_{j=1}^{p_C} \theta_{j,E}^C \psi_j^C \right).
\end{align*}
Finally, the co-variation structure between brain shape and connectivity, due to genetic factors, can be visualized by computing the surfaces and covariance functions, in their respective curved spaces, that are the Riemannian exponentials of the pairs of elements $(-c_1 v^S_{G},-c_2 v^C_{G})$ and $(c_1 v^S_{G},c_2 v^C_{G})$, with $c_1$ and $c_2$ appropriately chosen positive constants. Analogously, we can visualize the environmental co-variation structure between brain shape and connectivity, by computing the surfaces and covariance functions that are the Riemannian exponentials of the pairs of elements $(-c_1 v^S_{E},-c_2 v^C_{E})$ and $(c_1 v^S_{E},c_2 v^C_{E})$. The overall analysis is depicted in Figure~\ref{fig:analysis}.

\subsection{Brain shape modeling}\label{sec:model_shape}
We model shapes as deformations of the template $S_0$ and we define a Riemannian metric on the space of such deformations. In our application setting, this approach allows us to constrain our statistical estimates to be anatomically plausible shapes.

In detail, we introduce a diffeomorphic operator $\varphi$ mapping a sufficiently smooth Hilbert space $(\calV, \|\cdot\|_{\calV})$ onto a group $\calG$ of diffeomorphic functions. Formally, the space $\calV$ is the tangent space of $\calG$ at the identity map, and $\varphi$ is the associated exponential map. We define $\calG$ as follows. Let $\{v_t \in \calV: t \in [0,1]\}$ be a time-dependent $\calV$-valued process such that $\int_0^1 \| v_t \|^2_{\calV}\, dt< \infty$. Then, the solution $\phi_v:[0,1] \times \R^3 \ra \R^3$, at time $t=1$, of the Ordinary Differential Equation (ODE)
\begin{equation}\label{eq:ODE}
\frac{\partial \phi_v}{\partial t}(t,x) = v_t \circ \phi_v(t,x)\ \qquad t \in [0,1], x \in \R^3,
\end{equation}
with initial condition $\phi_v(0,x) = x$, is a smooth diffeomorphic deformation of $\R^3$ \citep[see, e.g.,][]{Younes2010}. The group $\calG$ consists of all such solutions of equation (\ref{eq:ODE}).

Equation (\ref{eq:ODE}) allows us to parameterize a diffeomorphic deformation $\phi_v(1,\cdot)$ (and therefore a shape $\phi_v(1,S_0)$ that preserves the topology of $S_0$) with a time-variant vector-field $\{v_t \in \calV: t \in [0,1]\}$. We then model the space of the time-variant vector-fields by defining $\{v_t:t \in [0,1]\}$ to be a time-variant vector field  which minimizes the quantity $\int_0^t \|v_t\|^2_\calV dt$, for a given initial vector field $v_0$ \citep{Miller2006}. Finally, the diffeomorphic operator is defined to be $\varphi_{v_0}(x) := \phi_v(1,x)$, where $v_0 \in \calV$ is the initial vector field generating $\{v_t:t \in [0,1]\}$, and $\phi_v$ is the solution of the ODE (\ref{eq:ODE}) for the computed $\{v_t:t \in [0,1]\}$.

With the notation of the previous section, we define $\gamma_i := \varphi_{v_i^S}$ with $i=1,\ldots,n$. The initial vector fields $(v_i^S)$ are estimated by minimizing a penalized mismatching functional, the details of which are left to Section~\ref{sec:estimation}. What is important from a statistical perspective is that a surface $S_i$, which belongs to a curved space, can now be represented by an element $v_i^S$ of the linear function space $\calV$, where $v_i^S$ is such that $S_i=\varphi_{v_i^S}(S_0)$.

\subsection{Brain connectivity modeling}\label{sec:model_cov}
The spatially normalized covariance functions $(C^0_i)$ associated with densely observed functional data on two-dimensional domains have the additional issue of being voluminous. Therefore, we first expand them into a finite functional basis $\{b_j\} \subset L^2(S_0)$, i.e.,
\[
	C^0_i(x,y) \approx \sum_{j=1}^{K} \sum_{l=1}^{K} C^K_{ijl} b_j(x) b_l(y),
\]
where $(C^K_i)_{jl}$ is the $K \times K$ covariance matrix that is a reduced representation of the covariance function $C^0_i$.

The functions $(b_j)$ are estimated from the data. A popular choice in neuroimaging is a set of indicator functions that are constant within connected regions of the cortical surface. This set of connected regions, also known as brain parcellation, defines functional sub-units on the cortical surface. An example of a parcellation is given by the color maps in the central panel of Figure~\ref{fig:data}. An alternative popular approach consists of estimating such a basis from an Independent Component Analysis of the fMRI data \citep{Calhoun2009}. Here we adopt the former approach.

Equipping the space of positive definite $K \times K$ covariance matrices with the $L^2$ distance results in variations around the mean that may not belong to the space of positive definite objects.  We therefore define a Riemannian metric on the space of symmetric positive-definite matrices by defining a smoothly varying scalar product on the tangent space, which is the linear space of $K \times K$ symmetric matrices. Such a Riemannian metric defines a geodesic distance on the space of covariance matrices that is given by the length of the shortest curve connecting any two covariances. \cite{Pennec2006} introduce an affine invariant Riemannian metric for positive-definite matrices that induces the distance $d_{\text{Riem}}(C_1, C_2) = \| \log(C_1^{-\nicefrac{1}{2}} C_2 C_1^{-\nicefrac{1}{2}}) \|_F$, where $C^{-\nicefrac{1}{2}} = V D^{-\nicefrac{1}{2}}V^T$, with $C = VDV^T$ denoting its spectral decomposition. A further option is the Cholesky distance  $d_{\text{chol}}(C_1, C_2) = \| \text{chol}(C_1) - \text{chol}(C_2)\|_F$, where $L = \text{chol}(C)$ denotes the Cholesky decomposition of a positive-definite matrix $C = L L^T$.  In \cite{Arsigny2006}, the log-Euclidean distance of two positive-definite matrices $C_1$ and $C_2$ is defined as $d_{\log}(C_1, C_2) = \| \log(S_1) - \log(S_2)\|_F$ , where $\| \cdot \|_F$ is the Frobenius norm and $\log(\cdot)$ the matrix logarithm, i.e., $\log(C) = V \log(D)V^T$, with $\log(D)$ denoting the diagonal matrix whose entries are the logarithms of the entries of $D$.

We model covariance matrices by equipping the space of covariances with the log-Euclidean metric defined in \cite{Arsigny2006}. For such a choice of the Riemannian metric, the geodesic distance is given by
\[
d(C_1,C_2) = \|\log(C_1) - \log(C_2) \|_F.
\]
The Fr\'echet mean $F \in \R^{K \times K}$ of $(C^K_i)$ is then defined as $F = \arginf_{M} \sum_{i=1}^n d(C^K_i,M)^2$. Given $C$, a $K \times K$ symmetric positive-definite matrix, and $L$ a $K \times K$ symmetric matrix, the Riemannian exponential and logarithmic maps have the form
\begin{align*}
\Exp_F(L) = \exp(\log(F) + \partial_L \log(F)), \qquad
\Log_F(C) = \partial_{\log(C) - \log(F)}\exp(\log(F)),
\end{align*}
where $\partial_L \log(F)$ and $\partial_{V}\exp(L)$ are respectively the differential of the matrix logarithm and the differential of the matrix exponential \citep{Arsigny2006}. The tangent representation $v_i^C$ of $C^K_i$ is then given by the $K \times K$ symmetric matrix $\Log_F(C^K_i)$. We also explored the application of other metrics with closed-form solution for the geodesic mean, such as the Cholesky metric and the square-root metric \citep{Dryden2009}. In our final application, for the chosen covariance size $K$, the different metrics did not seem to make a noticeable difference to the subsequent analysis. As mentioned previously, when $K$ is very large and potentially tends to infinity, the choice of metric might need to reflect this.

\section{Estimation}\label{sec:estimation}
\subsection*{Shape representation}
The surfaces $(S_i)$ need first to be registered; i.e., one-to-one correspondence needs to be established between subject-specific dense sets of landmarks $(x_l^{(i)}) \subset S_i$. In practice, the landmarks are the vertices of the triangulated surfaces approximating the idealized surface $S_i$. In order to avoid burdening the notation, we do not distinguish between idealized surfaces and associated triangulated surfaces, as this will be clear from the context.

The problem of image registration is common to any population analysis of images, however, the choice of the registration model generally depends on the specific application \citep{Zitova2003}. In our application setting, this step is performed by maximizing a measure of structural/functional `coherence' across subjects, while minimizing the amount of distortion introduced by the registration \citep{Fischl1999b,Yeo2010,Robinson2014,Robinson2018}.

Given the $n$ sets of registered landmarks, a template $S_0$ --- represented by the vertices $\{x_l^{(0)}\} \subset S_0$ --- is estimated by means of a Procrustes analysis. Such an analysis allows us to estimate the template while removing translation, size, and rigid rotations from the surfaces $(S_i)$. Then, the shape tangent space coordinates $({v}^S_i)$, associated with the surfaces $(S_i)$, are computed by solving a minimization problem of the form
\begin{equation}\label{eq:minimization_geo}
{v}^S_i = \argmin_{v_i \in \calV} \sum_l \| \varphi_{v_i}(x_l^{(0)}) - x_l^{(i)} \|^2_{\mathbb{R}^3} + \lambda \|v_i\|^2_\calV, \qquad i=1,\ldots,n,
\end{equation}
where the least-square term measures the similarity of the deformed template $\varphi_{v_i}(S_0)$ with $S_i$. The term $\|v_i\|^2_\calV$ can be intuitively understood as a regularizing term that measures the `energy' associated with the deformation. The constant $\lambda$ controls the trade-off between the empirical and regularizing term.

To obtain an unbiased estimate of the template $S_0$, with respect to the defined metric on the group of diffeomorphic functions $\calG$, we could update the template with ${S_0 \la \varphi_{\bar{v}}(S_0)}$, where $\bar{v}:= n^{-1} \sum_{i=1}^n {v}^S_i$. Subsequently, we could recompute $\{{v}^S_i\}$ by solving (\ref{eq:minimization_geo}) for the newly estimated template. These steps can then be iterated until convergence. Such a procedure is however prohibitive for computational reasons. Therefore, we fix the template to be the one resulting from the Procrustes analysis.

The space $\calV$ is modeled as a Reproducing Kernel Hilbert Space (RKHS) with a kernel that is a finite sum of Gaussian kernels of the type $K_\sigma(x,y) = \exp(-\frac{\|x-y\|^2_{\mathbb{R}^3}}{2 \sigma^2}) {I}_{3}$, for different choices of $\sigma > 0$. The minimization problem is approached with a BGFS optimization scheme \citep{Lewis2013}. We adopt the implementation in the MATLAB package \texttt{fshapetk} \citep{Charlier2015,Charlier2017a}.

\subsection*{Functional connectivity}
Given the $K \times K$ subject-specific covariance matrices $(C^K_i)$ (see Figure~\ref{fig:data} for an illustration), we first compute the least-square Fr\'echet mean estimate. Given our choice of the Riemannian metric, this has the closed-form solution \citep{Arsigny2006}
\[
F = \exp \left\{ \frac{1}{n} \sum_{i=1}^n \log{C^K_i} \right\}.
\]
The tangent space coordinates, in a matrix form, are given by $V_i^C = \Log_{F}(C^K_i)$. In practice, to circumvent stability issues related to the numerical computation of the differential of the matrix exponential and logarithm, we perform tangent expansion around the identity, i.e., work with the coordinates
\[
V_i^C = \log(C_i) - \log(F).
\]
Finally, the tangent space coordinates $v_i^C$, defined in Section~\ref{sec:model_cov}, are given by $v_i^C = \operatorname{vec}_{\operatorname{Sym}} \left( V_i^C \right)$, where $\operatorname{vec}_{\operatorname{Sym}}(L) = (l_{1,1}, \ldots, l_{K,K}, \sqrt{2} l_{1,2}, \ldots, \sqrt{2} l_{K-1,K})^T$ is a convenient vectorization operation for the space of symmetric matrices equipped with the Frobenius norm.

\subsection*{Mixed Effects Model}\label{sec:mixed_effect_impl}
We reformulate Model (\ref{eq:multi-trait}), by defining two $n \times (p_S+p_C)$ independent random matrices
\[
U \sim  \operatorname{MVN}(0_{n\times (p_S+p_C)}, I_n, I_{p_S+p_C}),\qquad V \sim  \operatorname{MVN}(0_{n\times (p_S+p_C)}, I_n, I_{p_S+p_C}).
\]
Moreover, we define $K_n^{\frac{1}{2}}, \Sigma_G^{\frac{1}{2}}$, and $\Sigma_E^{\frac{1}{2}}$ such that $K_n = K_n^{\frac{1}{2}} (K_n^{\frac{1}{2}})^T$, $\Sigma_G = \Sigma_G^{\frac{1}{2}} (\Sigma_G^{\frac{1}{2}})^T$, and $\Sigma_E = \Sigma_E^{\frac{1}{2}} (\Sigma_E^{\frac{1}{2}})^T$. Then, we can rewrite Model (\ref{eq:multi-trait}) as
\begin{align*}
&A \, | \, U, V =
XB + K_n^{\frac{1}{2}} U (\Sigma_G^{\frac{1}{2}})^T + V (\Sigma_E^{\frac{1}{2}})^T,
\end{align*}
thanks to the fact that
\[
K_n^{\frac{1}{2}} U (\Sigma_G^{\frac{1}{2}})^T \sim \operatorname{MVN}(0_{n\times (p_S+p_C)}, K_n, \Sigma_G), \qquad V (\Sigma_E^{\frac{1}{2}})^T \sim \operatorname{MVN}(0_{n\times (p_S+p_C)}, I_n, \Sigma_E).
\]

In practice, the kinship matrix $K_n$ is rank deficient, which is in fact a desirable property as it means that there are highly correlated samples that, intuitively, make it possible to disentangle the genetic and environmental covariance. When two samples are maximally correlated, as for instance in the case of monozygotic twins, the number of rows of $U$ can be reduced by one, and the two samples can be modeled with the same random effect.

The matrices $U$ and $V$ are then vectorized to a multivariate normal vector with independent samples, and the matrices $K^{\frac{1}{2}}$, $(\Sigma_G^{\frac{1}{2}})^T$, and $(\Sigma_E^{\frac{1}{2}})^T$ are reshaped accordingly. Finally, the entries of the unknown covariances $\Sigma_G$ and $\Sigma_E$ are estimated optimizing the REML criterion with respect to the parametrizing matrices $\Sigma_G^{\frac{1}{2}}$ and $\Sigma_E^{\frac{1}{2}}$. The proposed model has been implemented in \texttt{R} as a wrapper of the function \texttt{lmer} in the package \texttt{lme4} \citep{Bates2015}. In our application, $\Sigma_G$ and $\Sigma_E$ are unstructured covariance matrices. Nevertheless, the proposed model can be easily extended to handle structured covariance matrices $\Sigma_G$ and $\Sigma_E$.

\section{Statistical analysis \& Results}\label{sec:application}
\subsection*{Data \& Preprocessing}
This study focuses on the analysis of all 1003 healthy adult subjects, from the S1200 HCP data release \citep{VanEssen2013}, that have complete resting-state fMRI scans. The structural MRI images have been acquired at a resolution of 0.7mm isotropic and the resting-state fMRI images have been acquired at a spatial resolution of 2.0mm isotropic and a temporal resolution of 0.7s. Resting-state fMRI data were acquired in four runs of 15 minutes. During these fMRI scans, the subjects were not performing any explicit tasks. Further details on the acquisition process can be found in \cite{Glasser2013,Smith2013}. An extensive set of subject traits, such as behavioral and demographic covariates, are also provided. Moreover, the HCP dataset includes multiple members of the same families, resulting in a familial relatedness structure across samples.

The MRI and fMRI data have been pre-processed with the minimal pre-processing HCP pipeline \citep{Glasser2013}. In particular, white, pial, and midthickness surfaces of the cerebral cortex are reconstructed. We use the midthickness surfaces, which are surfaces that interpolate the midpoints between the white and pial surfaces, to describe the anatomy of the cerebral cortex and we refer to them as cortical surfaces. The four resting-state fMRI runs have been pre-processed to remove artifactual components in the data \citep{Smith2013}. The fMRI signals associated with neuronal activation on the cerebral cortex have been extracted and mapped onto the cortical surfaces, resulting in the data illustrated in Figure~\ref{fig:data}.

\subsection*{Spatial normalization}
The cortical surfaces are given in the form of two closed triangulated surfaces of 32K vertices, describing respectively the geometry of the left and right hemispheres. The 64K vertices have been brought in correspondence across subjects thanks to the application of a multi-modal surface alignment algorithm \citep{Robinson2014,Robinson2018}, which enables surface alignment based on both anatomical and functional features.

In the setting of joint shape and functional modeling, the importance of using the functional component of the data in the alignment procedure has been demonstrated, for instance, in \cite{Lila2019}. In the cited work, the authors propose a functional manifold surface alignment model embedded in the statistical analysis to improve the surface alignment. Here, we rely on the multi-modal spherical alignment of \cite{Robinson2018}, where extensive hyper-parameters tuning and validation have already been performed for the dataset in question.

In the next section, statistical shape analysis is performed on the cortical surfaces by treating the surface vertices as anatomical landmarks, given that these are in geometric and functional correspondence across subjects. Note that the defined surfaces cannot yet be regarded as \textit{shapes} \citep{Dryden2016} since non-physiological features, such as translation and rotations, are still present in the data.

\subsection*{Shape Analysis}
In order to project the surfaces in the shape-space --- which is, to remove Euclidean similarity transformations from the data --- we perform Generalized Procrustes Analysis \citep{Dryden2016} on the anatomical landmarks. This has the effect of removing translation, rigid rotation, and scale from the data, while iteratively estimating an average shape in the shape-space. Translation and rotation are discarded as these components do not have a physiological meaning. Scale, instead, is a positive scalar that does have a physiological meaning and its log-transformation $l_i$ will be incorporated in the final analysis. We denote with $S_i$ the brain surfaces projected onto the shape-space and with $S_0$ the estimated template average shape.

While features of the data that are non-descriptive of shape have now been removed, the landmark description of shapes does not guarantee that linear statistical models of the data, such as PCA, generate topologically valid shape trajectories. In our application, a valid shape trajectory is one that, for instance, does not lead to self-intersecting surfaces \citep[see, e.g.,][ for an illustration of the issue]{Vaillant2004}.

Therefore, we represent each surface $S_i$ with a vector field $v_i^S$ belonging to the linear space $\calV$. The vector field $v_i^S$ is computed by solving the minimization problem in (\ref{eq:minimization_geo}), leading to a function ${v}_i^S$ such that $S_i \approx \varphi_{{v}_i^S}(S_0)$, where $\varphi$ is the diffeomorphic deformation operator defined in Section~\ref{sec:model_shape}. In practice, two separate vector fields, one for each hemisphere, are estimated independently. We do not denote them separately as such a choice has computational advantages, but no other practical implications.

The RKHS space $\calV$ is defined by a kernel that is the sum of six isotropic Gaussian kernels in $\R^3$ with standard deviations $\{8, 4, 2, 1, 0.5, 0.1\}$. The penalty coefficient in (\ref{eq:minimization_geo}) is chosen to be $\lambda = 10^{-3}$. These hyper-parameters have been selected by experimenting with a small subset of the full cohort. Computations have been performed on a cluster where each node is equipped with an Intel Xeon E5-2650 2.2GHz 12-core processor with 96GB RAM and four Nvidia P100 GPUs. The minimization algorithm takes approximately 40 minutes for each subject and uses only one GPU.  The representing 1003 vector fields are computed in parallel on several nodes, greatly reducing the computation time needed.

We identify a set of demographic confounding variables (height, weight, sex, and age), which are demeaned, and their squares (when the confounder is a continuous variable) are included in the analysis. We then regress the confounders out of the RKHS coefficients representing $({v}_i^S)$ and the log-transformed size coefficients $(l_i)$ from the Procrustes Analysis.

We perform functional PCA analysis on $\{{v}_i^S\}$ leading to the representation ${v}_i^S \approx \bar{v}^S + \sum_{j=1}^{p_S}{A}_{i,j}^S {\psi}_j^S$. In contrast to the idealized basis expansion in Section~\ref{sec:model_shape}, we have a non-zero mean term $\bar{v}^S = n^{-1} \sum_{i=1}^n {v}_i^S$. In fact, due to the prohibitive computational costs incurred in computing ${v}_i^S$, we do not iteratively re-estimate the template until the mean term $\bar{v}^S$ becomes negligible, as noted in Section~\ref{sec:estimation}. We select the truncation level $p_S = 10$. This choice is mainly driven by computational limitations in the subsequent analysis. Our final shape representation of the cerebral cortex of the $i$th subject is given by $p_S$ scalar coefficients that are ${A}_{i,1}^S,\ldots,{A}_{i,p_S}^S$, and the log-transformed size coefficient $l_i$ from the Procrustes analysis.

\subsection*{Connectivity Analysis}
For each vertex of a surface $S_i$, we have four times series (one for each run) describing the resting-state neuronal activation of that location. Thanks to the anatomical and functional correspondence of the surface vertices across subjects, we can equivalently perform our analysis on the common template surface $S_0$.
We adopt an atlas of the template $S_0$ that assigns a label to each vertex of the template, defining a parcellation. For each run, and within each region of the parcellation, we compute a robust spatially averaged time-series that represents the brain activity of that entire region.

Many approaches have been proposed to define parcellation atlases \citep{Fischl2004,Desikan2006,Power2011,Yeo2011,VanEssen2012,Wig2014,Gordon2016,Glasser2016}. See \cite{Arslan2018} for a recent systematic review. Over the years they have tremendously improved in their granularity and ability to incorporate multi-modal imaging to define parcellations of the cortical surface. In this work, we rely on the popular Desikan-Killiany parcellation \citep{Desikan2006}, which defines $K=68$ cortical surface regions. More recent parcellations, as the one proposed in \cite{Glasser2016}, define up to $360$ regions. We opted for a courser parcellation to mitigate the effect of misregistration, which is the effect of small errors in the surface registration step.

For each subject, the $68 \times 4$ time-series are demeaned and variance normalized. A $68 \times 68$ covariance matrix is computed for each run and the average covariance across the four runs, $C_i^K$, is used to represent the $i$th subject functional connectivity. These covariance matrices are sometimes referred to as networks, or connectomes, as they quantify the functional connectivity between network nodes that are regions of the cortical surface.

We perform connectivity analysis in the log-Euclidean framework. As detailed in Section~\ref{sec:model_cov}, we compute the matrix logarithms of $C_i^K$ and compute the associated coefficients with respect to a Frobenius-orthogonal basis on the space of symmetric matrices. This leads to the computation of a set of representing tangent space coordinates $\{{v}_i^C\}$ that are vectors of dimension $2346$, which is the number of entries of the upper triangular part of the $68 \times 68$ covariance matrices.

In addition to those used in the shape analysis step, we identify additional confounding variables that are more directly related to the acquired fMRI signal \citep{Smith2015} (acquisition reconstruction software version, average subject head motion, systolic/diastolic blood pressure, hemoglobin A1C measured in blood, cube-root of total brain volume, and cube-root of total intracranial volume). The confounding variables are regressed out of the connectivity tangent coordinates $\{{v}_i^C\}$ (we do not rename the deconfounded tangent space coordinates).

We perform PCA on the deconfounded tangent space coordinates $\{{v}_i^C\}$, leading to the basis expansion ${v}_i^C \approx \bar{v}_i^C + \sum_{j=1}^{p_C} {A}_{i,j}^C {\psi}_j^C$. We truncate the basis expansion at $p_C = 10$. As in the shape analysis step, this choice is mainly driven by computational limitations in the subsequent joint shape/connectivity analysis. For the $i$th subject, connectivity is finally represented by 10 coefficients that are ${A}_{i,1}^C , \ldots, {A}_{i,p_C}^C$.

\subsection*{Family relatedness}
The statistical model proposed in Section~\ref{sec:model} relies on the presence of genetic relatedness among the subjects to disentangle genetic and environmental contributions to the covariation patterns between brain shape and connectivity. Of the 1003 subjects, 1001 had family relatedness information. The dataset consists of unrelated individuals, full siblings, half-siblings, dizygotic twins, and monozygotic twins. Specifically, there are 429 families, with a number of members that range from 1 to 6.

The relatedness matrix $K_n$, in the multivariable trait model (\ref{eq:multi-trait}), represents pairwise expected covariance, between subjects, that is due to familial relatedness. We estimate the matrix $K_n$ as $K_n = 2 \Phi_n$, with $\Phi_n$ the matrix of kinship coefficients \citep{Almasy1997, Lange2002}. We use Solar (\url{nitrc.org/projects/se_linux}) to compute the kinship matrix from the HCP family structure data. In larger population studies, an empirical genetic relatedness matrix could be computed from SNP data \cite[see, e.g.,][]{Kang2010,Wang2017b}.

\subsection*{Joint random-effects modeling}
In the previous steps of the analysis, for the $i$th subject, we have derived a vector of scalar variables
\begin{equation}\label{eq:descriptors}
{l}_i, {A}_{i,1}^S , \ldots, {A}_{i,p_S}^S, {A}_{i,1}^C , \ldots, {A}_{i,p_C}^C,
\end{equation}
with the first variable being a descriptor of brain size, the subsequent $p_S$ variables being descriptors of shape, and the final $p_C$ variables being descriptors of connectivity. The empirical $(1+p_S+p_C) \times (1+p_S+p_C)$ covariance matrix $\Sigma$, computed from these variables, describes the pairwise first-order dependencies between such descriptors, hence the co-variation structure between shape and connectivity that is due to both genetic and environmental contributions. We instead want to leverage the familial relatedness matrix $K_n$ and apply the joint mixed model in Section~\ref{sec:model} to disentangle the additive covariance components $\Sigma_G$ and $\Sigma_E$, which are respectively due to genetic and environmental contributions.

\begin{figure}[!htb]
\centering
\includegraphics[width=0.8\textwidth]{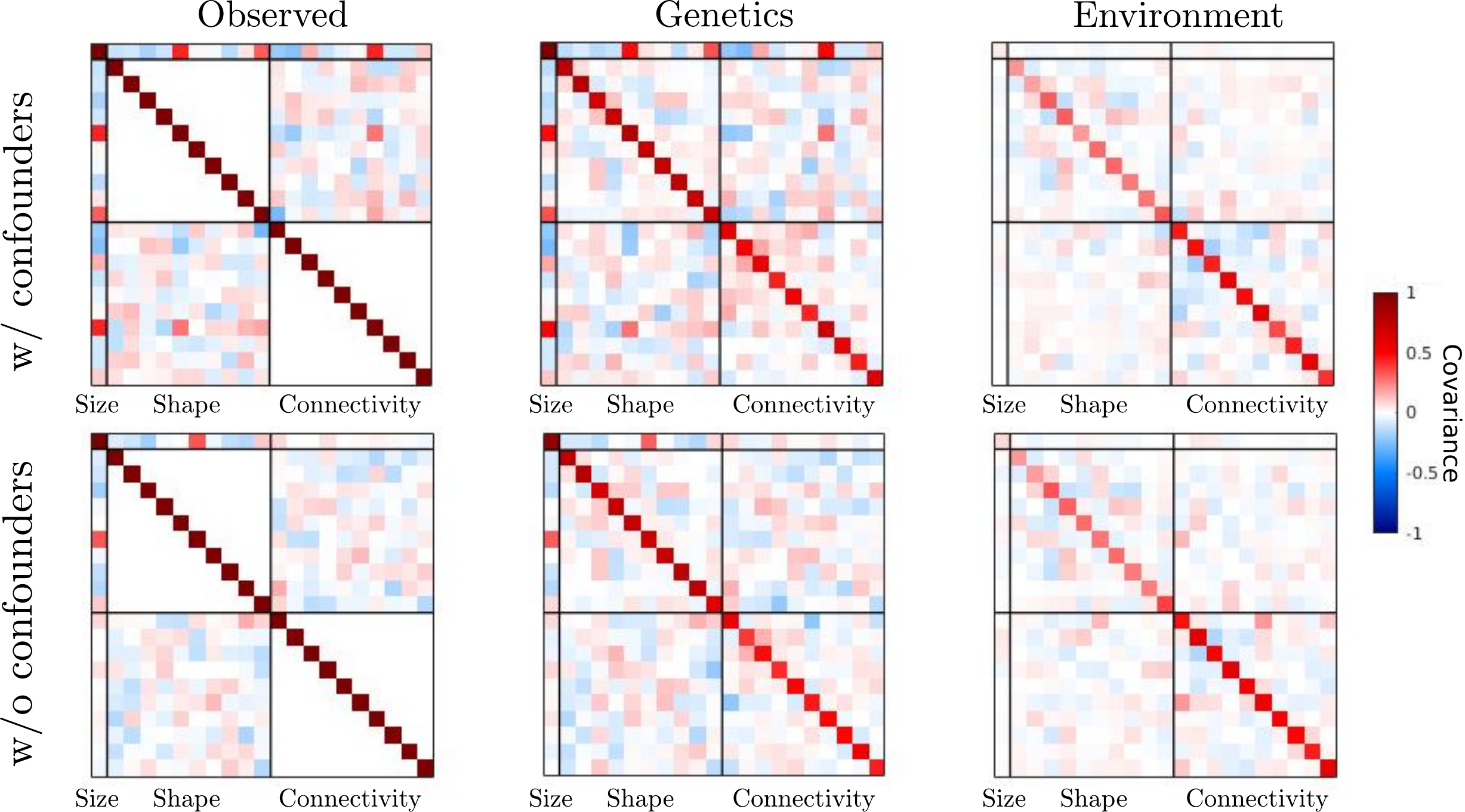}
\caption[]{On the left panel, we plot the empirical covariance of the $n \times (1+p_S+p_C)$ data matrix of size, shape, and connectivity descriptors. Each descriptor (i.e., each column of the data matrix) has been normalized to have unit standard deviation. On the middle and right panel, we can find the latent covariance components that are due to genetic and environmental factors, as recovered by the variance component model proposed.}
\label{fig:RE}
\end{figure}

The covariances $\Sigma_G$ and $\Sigma_E$ are estimated by optimizing the REML criterion associated with Model (\ref{eq:multi-trait}), without a fixed effects term. For our choice of the truncation levels ($p_S = p_C = 10$), the minimization of the REML takes approximately 8 hours. This is the limiting factor in the choice of the truncation levels. Nonetheless, in general, care should be taken in increasing $p_S$ and $p_C$, as the descriptors could start capturing smaller scale variations in shape and connectivity driven by small and unavoidable errors in the surface alignment steps \citep[for an illustration of the issue, see Section S3 in the supplementary material for][]{Lila2019}.

\subsection*{Results}
In Figure~\ref{fig:RE}, we show respectively the estimates of $\Sigma$ (the empirical covariance structure), $\Sigma_G$ (the covariance structure due to genetic contributions), and $\Sigma_E$ (the covariance structure due to unique environmental contributions) of the $1+p_S+p_C$ descriptors of size, shape, and connectivity.

We measure the heritability, i.e., the portion of shape and connectivity variability explained by additive genetic contributions as
\[
h^2 = \frac{\operatorname{trace} (\Sigma_G)}{\operatorname{trace} (\Sigma_G + \Sigma_E)}.
\]
Our overall heritability estimate is $h^2 = 0.61$, which means we estimate that 61\% of the source of variability in the data (post-PCA) is due to genetic contributions.
However, if we measure the heritability of size, shape, and connectivity separately, we obtain the estimates 0.92, 0.71, and 0.47, respectively. These are consistent with recent estimates in the literature \citep{Barber2021,Pizzagalli2020}, and with the intuition that functional features are more easily affected by life events and environmental factors, and therefore are less heritable. This is also clear from the diagonal entries of $\Sigma_G$ and $\Sigma_E$ in Figure~\ref{fig:RE}.

\begin{figure}[!htb]
\centering
\includegraphics[width=0.8\textwidth]{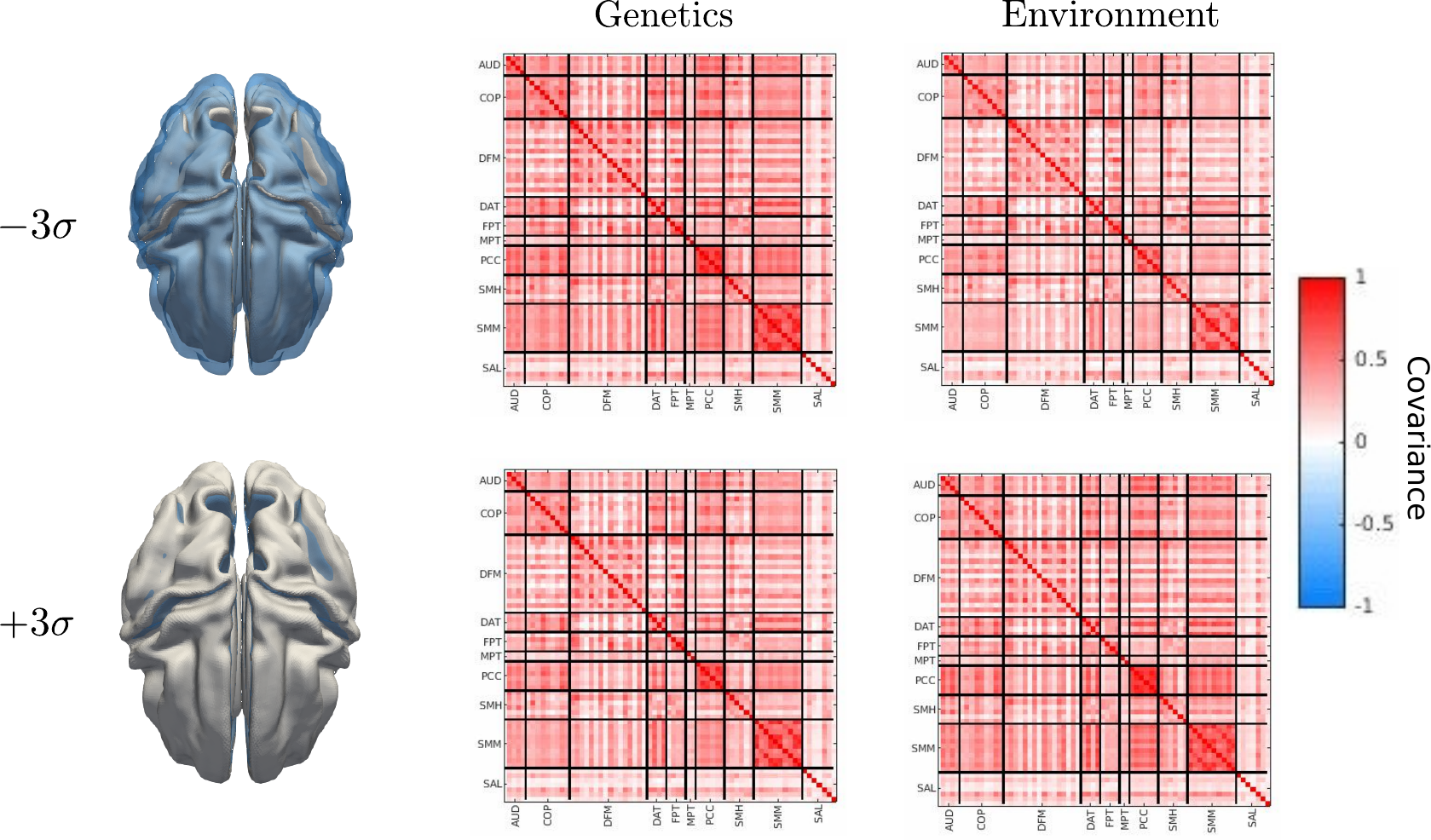}
\caption{An illustration of the changes in connectivity levels associated with a $\pm 3 \sigma$ variation in the univariate variable log-transformed brain size, where $\sigma$ is the standard deviation of such a variable. The corresponding anatomical objects are shown as gray-colored surfaces. The blue surfaces represent the fixed average brain and are shown in order to have a fixed reference between plots and be able to appreciate the differences between the two gray surfaces. In the areas where only the gray surface is visible, this is overlaying the blue surface. The associated changes in genetic and environmental connectivity levels, shown in the form of covariance matrices, are minimal. Instead, running the analysis  without removing the effect of confounders displayed a general increase in connectivity that is associated with an increase in brain size. In order to improve the readability of the covariance plots, we have clustered their nodes in pre-identified communities \citep{Power2011}: auditory network (AUD), cingulo-opercular network (COP), default mode network (DMN), dorsal attention network (DAT),  frontoparietal network (FPT), medial parietal network (MPT), somatosensory/motor network (SMH \& SMM), and salience network (SAL).}
\label{fig:reg}
\end{figure}

\begin{figure}[!htb]
\includegraphics[width=\textwidth]{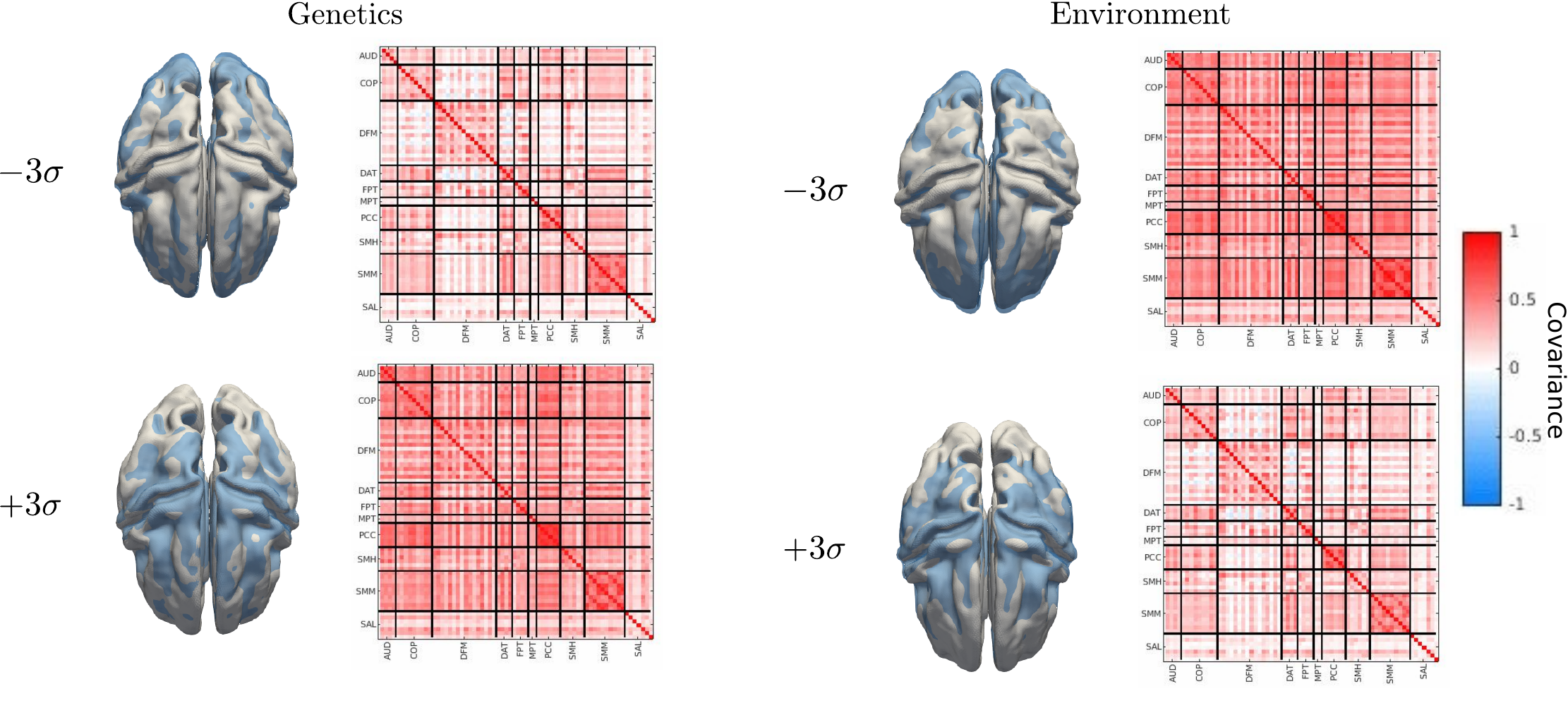}
\caption{Illustration of the shape and connectivity CCA main modes of co-variation that are due to genetic and environmental contributions. Specifically, we display $\pm 3 \sigma$ changes in shape (gray-colored surfaces) and connectivity that are most correlated according to the estimated covariance structure $\Sigma_G$ (left panel) and $\Sigma_E$ (right panel). The variable $\sigma$ here denotes the corresponding standard deviation of the unnormalized shape and connectivity descriptors in equation (\ref{eq:descriptors}), once these have been projected along the directions representing the respective CCA modes of variation. Note that $\sigma$ has different values for shape/connectivity and for genetic/environmental components. As in Figure~\ref{fig:reg}, the blue surfaces are shown in order to have a fixed reference across the four panels and be able to appreciate the differences between the $\pm 3 \sigma$ gray surfaces. The main modes of co-variation display an association between a global change in shape and a global change in connectivity levels, with a contrasting behavior in the genetic and environmental components. Larger variations in connectivity levels are displayed between functional communities rather than within communities.}
\label{fig:cca}
\end{figure}

\begin{figure}[!htb]
\centering
\includegraphics[width=0.5\textwidth]{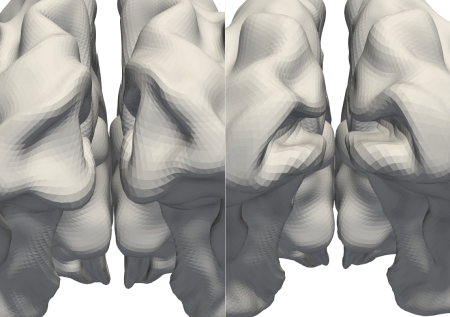}
\caption{On the left, the shape configuration identified by the tangent vector $- 3\sigma v^S_{G}$. On the right, the shape configuration identified by the tangent vector $+3\sigma v^S_{G}$. The tangent vector $v^S_{G}$ represents shape changes, due to genetic contributions, in the main mode of co-variation between shape and connectivity. These are effectively a different view of the same gray surfaces shown on the top-left and bottom-left panels of Figure~\ref{fig:cca}.
The figure highlights the ability of the proposed framework to capture non-trivial localized variations in the brain shape, such as the formation of a sulcus.}\label{fig:cca_zoom}
\end{figure}

These heritability estimates are a byproduct of our analysis. The main contribution of our work is instead to quantify the dependence structure between anatomical and functional features and display this dependence in terms of actual variations of the neurobiological object considered. In Figure~\ref{fig:reg}, we show the results of a linear regression model, between brain size and connectivity, formulated using the estimated genetic and environmental covariances. After the linear model is fitted, we display the variation in connectivity associated with a $\pm 3 \sigma$ variation of the log-transformed size variable, where $\sigma$ is its standard deviation. The results do not display large associated variations in connectivity. However, when running the analysis without removing the confounders from the size and connectivity descriptors, we observe that a general increase in connectivity is associated with an increase in brain size, both in the genetic and environmental components. This association is driven by the confounding factors.
\allowbreak
In Figure~\ref{fig:cca}, we display the results of the CCA between shape descriptors and connectivity descriptors. Specifically, we compute the tangent coordinates $(-3 \sigma^S_G\, v^S_{G},-3 \sigma^C_G\, v^C_{G})$ and $(3 \sigma^S_G\, v^S_{G},3 \sigma^S_G\, v^C_{G})$, for the genetic contribution, and $(-3 \sigma^S_E\, v^S_{E},-3 \sigma^C_E\, v^C_{E})$ and $\allowbreak (3 \sigma^S_E\, v^S_{E},3 \sigma^C_E\, v^C_{E})$, for the environmental contribution, as detailed in Section~\ref{sec:model}. The variables $\sigma^S_G$,$\sigma^C_G$,$\sigma^S_E$ and $\sigma^C_E$ denote the standard deviation of the unnormalized shape and connectivity descriptors along the associated mode of co-variation. With a slight abuse of notation, in Figure~\ref{fig:cca}-\ref{fig:cca_zoom}, we denote these standard deviations by $\sigma$.

We then plot the shape and connectivity configurations identified by these tangent coordinates. Our main modes of genetic and environmental co-variation highlight an association between a global variation in brain shape and a global change in brain functional connectivity levels that seem to show a contrasting behavior in the genetic and environmental modes of co-variation.

The proposed approach to modeling shape and connectivity allows us to capture both global and local variations. A more meticulous exploration of the results demonstrates a clear advantage of the adopted approach to shape modeling. In particular, in Figure~\ref{fig:cca_zoom}, we show local shape changes that are associated with the mode of variation $v^S_{G}$. We can see non-trivial shape variations involving the formation of a sulcus. Capturing such fine-grained variations would in general not be possible with the simple shape descriptors (e.g., surface area) classically adopted in the neuroimaging literature.


\section{Discussion}\label{sec:conclusion}
In this work, we propose a statistical Riemannian approach for the analysis of samples that are brain shapes and brain connectivity.  In the proposed framework, we embed a variance component model that exploits family relatedness structure among samples to separate brain shape and connectivity co-variation that is due to genetic and environmental factors.

The proposed Riemannian modeling approach allows us to estimate trajectories in the spaces of shapes and connectivity that are constrained to belong to their respective non-Euclidean spaces of anatomically/physiologically meaningful estimates. Specifically, we are able to exclude shapes that are not topologically equivalent to the shapes in the sample, e.g., self-intersecting shapes, and we are able to exclude functional connectivity estimates that are not symmetric positive-definite objects. Moreover, the shape modeling approach proposed in this paper can also be easily extended to incorporate heterogeneous types of imaging data, such as volumetric representations of subcortical brain structures and bundles of axons estimated from Diffusion Tensor Imaging \citep{Feydy2017}. The proposed framework can be readily applied to the analysis of different anatomical objects.

When it comes to shapes and covariances, different representation models and metrics have been proposed in the literature. The choices we have made in this work are mainly driven by the reasons aforementioned. However, the exploration of different shape metrics, such as the Elastic Metric \citep{Kurtek2011,Jermyn2012,Jermyn2017}, is also a promising direction for future work. A particularly attractive property of the latter is the ability to integrate the registration step with the computation of the representation. Nonetheless, it is in principle more complicated to enforce topological constraints.

One limitation of the proposed variance component model is that it is based on a reduced dimension representation of shape and connectivity. It is, therefore, of interest to extend the current approach to work directly on the bivariate functional tangent space representations of shape and connectivity. Nevertheless, due to the high-dimensionality of the data, this is currently prohibitive, not only from a computational perspective but also due to the need to incorporate regularizing penalties to control for the nearly co-linear modes of co-variation that arise when estimating canonical correlation components from functional data. Further, in our analysis, we only focus on two factors contributing to the covariance structure of the traits: additive genetic and unique environmental factors. It is of course of interest to extend the proposed model to other factors, such as gene-environment interactions. It is also equally important to apply the proposed framework to the large imaging datasets nowadays available, such as the UK Biobank \citep{Sudlow2015}, while exploiting the genomic component of these datasets to estimate genetic relatedness. These datasets are more representative of the general population and could allow us to answer questions related to the effect of genes, age, and diseases on anatomical and functional properties of the brain.


\appendix
\appendixpage

\section{Simulations}
In this section, we perform simulations to assess the finite sample estimation properties of the variance component model (\ref{eq:multi-trait}).

\begin{figure}[!htb]
\centering
\includegraphics[width=.6\textwidth]{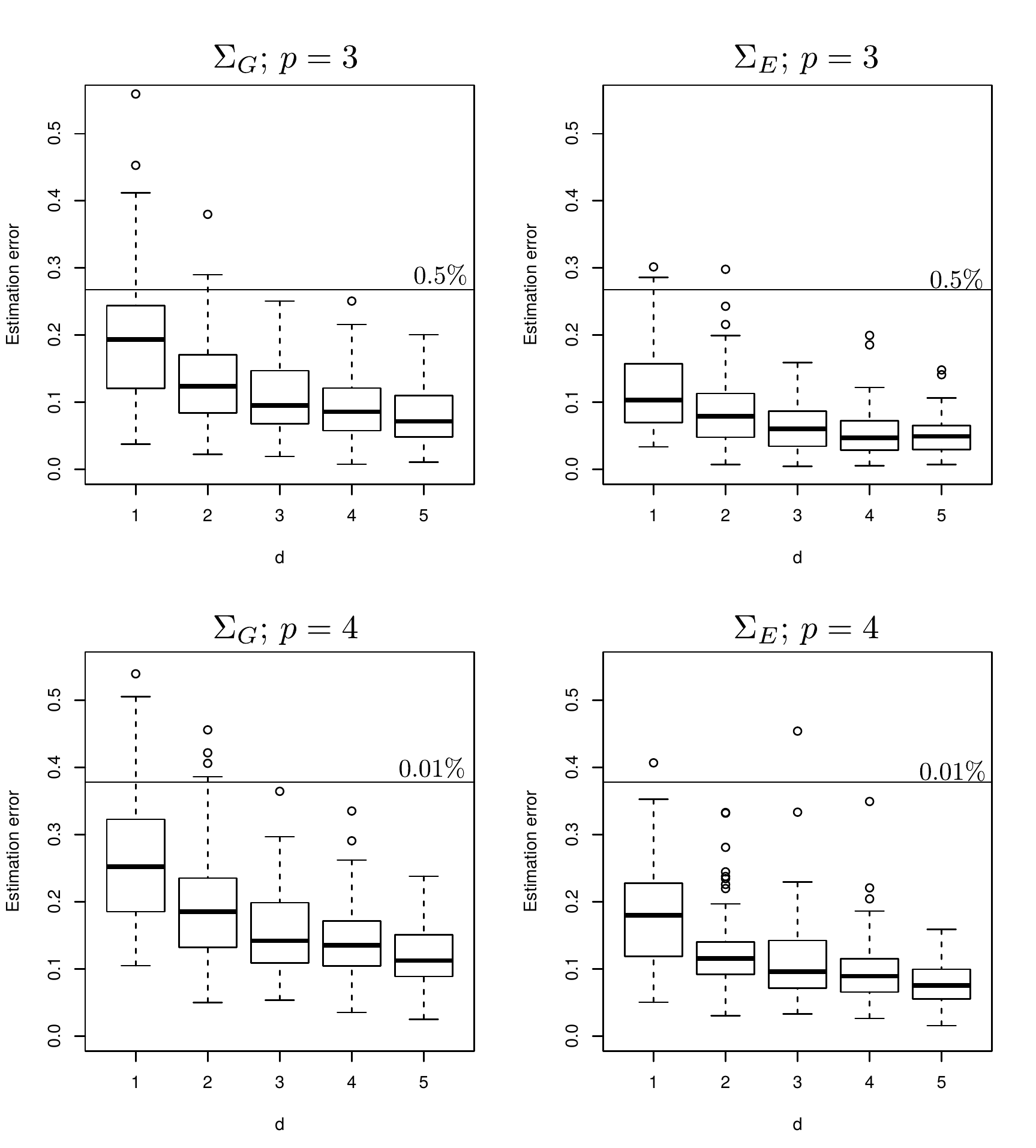}
\caption{Boxplots describing the performances of the variance component model, in estimating $\Sigma_G$ and $\Sigma_E$, as a function of the variable $d=1,\ldots,5$, which is a multiplicative factor in the sample size $nd$. The estimation error is measured with the Frobenius norm of the difference between the covariance to be estimated and its estimate. We run 100 simulations for each choice of $d$ and $p$. The horizontal lines denote the 0.5 and 0.01 percentiles, respectively for $p=3$ and $p=4$, of the smallest estimation errors of a random guess estimator.}
\label{fig:boxplot}
\end{figure}

We generate two $p \times p$ correlation matrices $\Sigma_G$ and $\Sigma_E$ as independent samples of a random correlation matrix which has a uniform distribution over the space of positive-definite correlation matrices \citep{Joe2006}. In order to obtain a simulation setting that is most similar to that of our application, we use the $n \times n$ kinship matrix $K_n$ from the HCP dataset, with $n = 1001$. Nevertheless, to study the effect of a hypothetical increase in sample size on our estimates, we introduce the $nd \times nd$ kinship matrix $I_d \otimes K_n$. Such a kinship matrix describes a situation where $nd$ samples are instead available, with $d$ unrelated groups of $n$ samples that have correlation structure represented by $K_n$.

We then generate our data following Model (\ref{eq:multi-trait}), i.e.,
\begin{equation*}
\begin{aligned}
&A \,| \,G,E  =
G + E\\
&G \sim \operatorname{MVN}(0_{nd\times p}, I_d \otimes K_n, \Sigma_G),\\
&E \sim \operatorname{MVN}(0_{nd\times p}, I_{nd}, \Sigma_E),
\end{aligned}
\end{equation*}
where no fixed effects is included to reflect the setting of our final application. The $nd \times p$ matrix $A$ represents a set of simulated tangent space coordinates.

For every choice of $d=1,\ldots,5$, and $p=3,4$, we generate 100 datasets and then apply the mixed effects model in Section~\ref{sec:mixed_effect_impl} to compute the estimates $\hat{\Sigma}_G$ and $\hat{\Sigma}_E$ of $\Sigma_G$ and $\Sigma_E$. We measure the accuracy of the estimate of the genetic component as $\| \hat{\Sigma}_G-\Sigma_G\|_F$ and that of the environmental component as $\| \hat{\Sigma}_E-\Sigma_E\|_F$, where $\|\cdot\|_F$ is the Frobenius norm.

We show the results of the simulations in Figure~\ref{fig:boxplot}, where we can see that, as expected, an increase in sample size results in a lower estimation error. Moreover, as in the data generation model, we generate $1000$ correlation matrices representing correlation matrices to be estimated. For each of the $1000$ correlation matrices, we generate $1000$ associated naive estimates, which are correlation matrices from the same distribution, and compute their Frobenius distance from the true correlation. The empirical distribution of the computed Frobenius norms describes the performance of the naive random estimator. We then select the $0.5$ percentile, for $p=3$, and $0.01$ percentile, for $p=4$, of the Frobenius norms. These are the horizontal lines in Figure~\ref{fig:boxplot}. We can see that most of the estimates from the mixed-effects model are well below the selected threshold.
\bibliographystyle{abbrvnat_brief}

\bibliography{Bibliography} 

\end{document}